# Ultrafast magneto-photocurrents as probe of anisotropy relaxation in GaAs

Christian B. Schmidt,[1*] Shekhar Priyadarshi,[1,2] Klaus Pierz,[1] and Mark Bieler[1]

[1] Physikalisch-Technische Bundesanstalt, Bundesallee 100, 38116 Braunschweig, Germany

[2] Institute of Physical Chemistry, University of Hamburg, 20146 Hamburg, Germany

We induce ultrafast photocurrents in a GaAs crystal exposed to a magnetic field by optical femtosecond excitation. The magneto-photocurrents are studied by time-resolved detection of the simultaneously emitted THz radiation. We find that their dynamics differ considerably from the dynamics of other photocurrents which are expected to follow the temporal shape of the optical intensity. We attribute this difference to the influence of carrier-anisotropy relaxation on the magneto-photocurrents. Our measurements show that the anisotropy relaxation for carrier densities ranging between $10^{16}\,\mathrm{cm}^{-3}$ and $5\times 10^{17}\,\mathrm{cm}^{-3}$ occurs on two different time scales. While the slow time constant is approximately 100 fs long and most likely governed by electron-phonon scattering, the fast time constant is on the order of 10 fs and presumably linked to the valence band. Our studies not only help to better understand the microscopic origins of optically induced currents but – being even more important – show that magneto-photocurrents can be employed as novel probe of anisotropy relaxation in GaAs. This technique is applicable to all non-centrosymmetric bulk semiconductors.

---

* Electronic mail: christian.b.schmidt@ptb.de

Optical excitation of non-centrosymmetric semiconductors creates carrier distributions with a certain shape in momentum space. This process is typically referred to as optical alignment of carriers.[1,2] In particular, femtosecond laser excitation of asymmetric carrier distributions $\rho_{11}$ with $\rho_{11}(-k_x) \neq \rho_{11}(+k_x)$ has been intensively investigated in the past two decades.[3–10] Such excitation processes allow ultrafast optical coherent control of spin-polarized charge currents[3,7,10,11] and pure spin currents[12–14] possibly enabling future spintronic applications.[15–19]

In addition to asymmetric carrier distributions, optical alignment may also lead to anisotropic distributions with $\rho_{11}(k_x) \neq \rho_{11}(k_y)$.[1,20] Yet applications utilizing anisotropic carrier distributions are very rare. A certain type of magneto-photocurrent[2,21–24] is one among few examples in which the anisotropic carrier distribution leads to a macroscopic response. Since this current occurs at the semiconductor surface and requires a magnetic field and linearly polarized optical excitation, we refer to it as linear surface magneto-photocurrent ($j_{\text{LSMC}}$). Recently we have shown that the $j_{\text{LSMC}}$ can be induced in (001)-oriented bulk GaAs on ultrafast time scales and separated from other current contributions by symmetry considerations.[21]

Here we induce the $j_{\text{LSMC}}$ in (110)-oriented bulk GaAs via femtosecond optical excitation and study the simultaneously emitted time-domain THz traces. We develop a scheme in which we express the $j_{\text{LSMC}}$ by other photocurrents and a term denoting anisotropy relaxation. This enables us to accurately model the $j_{\text{LSMC}}$ and to determine anisotropy relaxation times in GaAs. Our results not only help to understand better the temporal dynamics of magneto photocurrents but also suggest that the $j_{\text{LSMC}}$ can be employed as novel intrinsic probe of anisotropy relaxation in semiconductors as compared to previously employed external optical probes.[25,26] This is also important from a fundamental point of view since the decay of the anisotropy is expected to occur on a different time scale as compared to the momentum scattering or dephasing[20] but corresponding experimental studies do not exist.

We first introduce the $j_{\text{LSMC}}$ and theoretically compare it to other photocurrents. The $j_{\text{LSMC}}$ occurs due to imperfect surface reflections of an anisotropic carrier distribution which creates an asymmetry in momentum space.[23] Our time-dependent model is based on the time-independent model of Alperovich et al.[23,27] and the corresponding process is pictured in Fig. 1(a) exemplarily for electrons. In the left-hand-side picture we show electrons which are created by an optical excitation pulse with an electric field envelope $E_{\text{opt}}(t)$ at a certain time $t_0$ and certain position in the sample. For illustration purposes we neglect electrons created at other positions in the sample. The corresponding generation term has the time dependence[8,28] $g(t) = E_{\text{opt}}(t)\{E_{\text{opt}}(t) * H(t)\exp(-t/\tau_d)\}$, with $H(t)$ being the Heaviside function; $\tau_d$ is the dephasing time and $*$ the convolution operator. For zincblende structures and linearly polarized light $g(t)$ expresses an anisotropic electron distribution in momentum space with the **k** dependence being implicit. For electron/heavy-hole transitions, the distribution has the form of a torus with the axis of revolution being parallel to the polarization direction of $E_{\text{opt}}(t)$.[1] An in-plane magnetic field now turns $g(t)$ around the magnetic field vector while its anisotropy simultaneously relaxes. Thus at $t_1 > t_0$ the electron distribution is modified to $[g(t)] * [H(t)\sin(\omega_c t)\exp(-t/\tau_a)]$, where $\omega_c$ is the cyclotron frequency and $\tau_a$ the anisotropy relaxation time. This distribution will still be anisotropic if $t_1 \lesssim \tau_a$. If the electrons hit the surface at $t_2 > t_1$, they will be scattered elastically and inelastically. While elastic scattering only inverts the component of the carrier momentum normal to the surface, inelastic scattering leads to changes of the in-plane momentum. Only for anisotropic carrier distributions this inelastic scattering will induce a net momentum parallel to the surface constituting the $j_{\text{LSMC}}$.[27] The $j_{\text{LSMC}}$ decays with the momentum relaxation time $\tau_m$, yielding

$$j_{\text{LSMC}}(t) \propto g(t) * \left[H(t)\sin(\omega_c t)\exp\left(-t/\tau_a\right)\right] * \left[H(t)\exp\left(-t/\tau_m\right)\right]. \qquad (1)$$

In the experiments described below we will induce the $j_{\text{LSMC}}$ by excitation with femtosecond laser pulses and detect it by time-resolved measurement of the simultaneously emitted THz radiation, which is proportional to the time derivative of the current. Unfortunately, the detected THz trace is distorted by propagation and detection effects [29,30] leading to a band-pass filtering of the THz signal. This makes it very difficult, if not impossible, to accurately determine the time constants of Eq. (1) just from the measured THz traces of the $j_{\text{LSMC}}$ alone.

To circumvent this problem, we induce other photocurrents under the same experimental conditions and express the $j_{\text{LSMC}}$ using these photocurrents. The shift current[4,8,31] arises due to the spatial shift of the electron charge during optical interband excitation from the valence to the conduction band. The injection current[4,8] occurs due to an asymmetric carrier distribution induced by optical excitation. Phenomenologically, the time dependence of both currents is given by[8,28]:

$$j_{\text{shift}}(t) \propto g(t), \qquad (2)$$

$$j_{\text{inj}}(t) \propto g(t) * \left[H(t)\exp(-t/\tau_m)\right]. \tag{3}$$

Comparing Eqs. (1), (2), and (3) we see that $j_{\text{LSMC}}$ can be expressed by $j_{\text{inj}}$ or $j_{\text{shift}}$. Consequently, we can also express the measured THz traces $e_{\text{LSMC}}$ by $e_{\text{inj}}$ or $e_{\text{shift}}$ using the same expressions. The study of both $j_{\text{inj}}$ and $j_{\text{shift}}$ is necessary since $j_{\text{inj}}$ and $j_{\text{LSMC}}$ do not exist in the same sample as described in the next paragraph.

As samples we used (110)-oriented bulk GaAs and (110)-oriented GaAs/Al$_{0.3}$Ga$_{0.7}$As multiple quantum wells (QWs) with a well width of 8 nm. While the $j_{\text{LSMC}}$ only occurs in the bulk sample, $j_{\text{inj}}(t)$ exists only in the QW sample and $j_{\text{shift}}$ exists in both samples.[2,8,32] In the following discussion we take the $\hat{x}$, $\hat{y}$, and $\hat{z}$ axes as the [001], [1$\bar{1}$0], and [110] directions, respectively, for both samples. For optical excitation we used 180 fs long laser pulses with a spectral full width half maximum (FWHM) of 10 meV focused down to a spot size of 220 μm FWHM at normal incidence. The temporal shape of the THz traces is measured using electro-optical sampling (EOS) in transmission geometry. The setup is the same as in Ref. [21].

In the bulk sample we measured $j_{\text{LSMC}}$ and $j_{\text{shift}}$ along the $\hat{x}$ direction using linearly polarized optical excitation along the [$\bar{1}$11] and [1$\bar{1}$1] directions and a magnetic field of 1.25 T along the $\hat{x}$ direction. Using this symmetry $j_{\text{shift}}$ is independent on the two specified optical polarizations and the magnetic field while $j_{\text{LSMC}}$ and therefore also $e_{\text{LSMC}}$ reverse its sign upon change of the polarization direction from [$\bar{1}$11] to [1$\bar{1}$1] or by inverting the magnetic field; see Fig. 1(b). Thus, by adding and subtracting THz traces measured for different optical polarizations and magnetic fields, we obtained $e_{\text{shift}}$ and $e_{\text{LSMC}}$, respectively. In the QW sample we measured $e_{\text{shift}}$ and $e_{\text{inj}}$ along the $\hat{y}$ direction employing linearly polarized excitation along the [$\bar{1}$11] and [1$\bar{1}$1] directions and left- and right-handed circularly polarized excitation, respectively. In both cases the currents reverse their sign upon change of the polarization state; see Fig. 1 (c). Thus, by subtracting the corresponding THz traces we obtained $e_{\text{inj}}$ and $e_{\text{shift}}$. It should be emphasized that other optically induced currents, such as surface currents rotated in the magnetic field and flowing along the $\hat{y}$ direction in the bulk sample, were suppressed by THz polarizers and the THz-polarization dependent EOS.[21] Moreover, another type of magneto-photocurrent which simultaneously occurs with the $j_{\text{LSMC}}$ in (001)-oriented GaAs[21] does not exist in our specific geometry.

THz traces of all four currents ($e_{\text{inj}}$, $e_{\text{shift,QW}}$, $e_{\text{shift,bulk}}$, $e_{\text{LSMC}}$) are plotted in Fig. 2(a) for an excitation photon energy of 1.52 eV with an average pump power of 1 W (125 MWcm$^{-2}$ peak intensity), corresponding to a carrier density of approximately $5 \times 10^{17}$cm$^{-3}$ in the bulk and $9 \times 10^{11}$cm$^{-2}$ in the QW sample. It is obvious that the shapes of $e_{\text{inj}}$, $e_{\text{shift,QW}}$, $e_{\text{shift,bulk}}$ are nearly identical. This leads to the conclusions that $j_{\text{shift,QW}}$ is identical to $j_{\text{shift,bulk}}$[4] and that momentum relaxation is too fast to be observed in our measurements and, thus, can be neglected; compare Eqs. (2) and (3). In contrast, $e_{\text{LSMC}}$ has a completely different shape. First off all the THz peak is shifted to later times and the second minima is considerably reduced indicating much slower dynamics of $j_{\text{LSMC}}$ compared to the other currents. To further illustrate the difference between $j_{\text{LSMC}}$ and the other currents, we plotted the temporal shift of the THz peak versus optical pump power for each current in Fig. 2(b). The shifts of $e_{\text{shift,QW}}$, $e_{\text{shift,bulk}}$, and $e_{\text{inj}}$ are very similar while the shift of $e_{\text{LSMC}}$ is approximately three times larger.

To investigate the different shape of the THz traces in more detail, we rewrite Eq. (1) as:

$$e_{\text{LSMC,fit}}(t) \propto e_{\text{shift}}(t) * \sum_i a_i H(t) \omega_{c,i}\, t \exp\left(-t/\tau_{a,i}\right), \qquad (4)$$

where we have replaced $g(t)$ by $e_{\text{shift}}(t)$ using Eq. (2) and neglected the term $H(t)\exp(-t/\tau_m)$ since Fig. 2(a) demonstrates that it does not influence our measurements. (Momentum relaxation in the QW sample is too fast to be detected and is expected to be even faster in the bulk sample due to the higher degree of freedom for scattering processes.)[33] Moreover we approximate $\sin(\omega_c t)$ with $\omega_c t$ since we do not detect any change of the shape of the THz traces versus magnetic field, but only a linear dependence of its amplitude.[21] This is equal to the condition $1/\tau_a \gg \omega_c$, with $\omega_c/2\pi$ being approximately equal to 550 GHz and 70 GHz for electrons and heavy holes, respectively, at 1.25 T. Finally the sum over $i$ takes into account that that different bands (i.e., electrons and holes) contribute to the $j_{\text{LSMC}}$[23] and the factor $a_i$ denotes the strength of the individual contributions.

With help of Eq. (4) we can express $e_{\text{LSMC}}(t)$ using the measured $e_{\text{shift}}(t)$ and a certain number of fit parameters. In the following discussion we will first show that we indeed need a double-exponentially decaying function, i.e., $i = 1,2$ in Eq. (4); a single-exponentially decaying function does not provide very good fit results. This is shown in Fig. 3(a) where we plot $e_{\text{LSMC}}$, $e_{\text{shift}}$, and $e_{\text{LSMC,fit}}^{\text{single exp.}}$, with the latter being obtained from Eq. (4) for a single-exponentially decaying function. Although the peak shift between $e_{\text{shift}}$ and $e_{\text{LSMC}}$ is nicely reproduced by $e_{\text{LSMC,fit}}^{\text{single exp.}}$, considerable discrepancies between the shapes of $e_{\text{LSMC}}$ and $e_{\text{LSMC,fit}}^{\text{single exp.}}$ are obtained. The least squares error between $e_{\text{LSMC}}$ and $e_{\text{LSMC,fit}}^{\text{single exp.}}$ are indicated in Fig. 3(b) as horizontal lines for two different carrier densities. In Fig. 3(b) are also shown the least squares errors obtained from a fit with a double-exponentially decaying function keeping the fast relaxation time fixed. For very small values of the fixed relaxation time the least squares error can be considerably reduced. We take this as a strong indication that the anisotropy decays with a double-exponential function. Yet, the exact determination of the fast time constant of this function is not possible, since a corresponding fit does not converge. This is also visualized in Fig. 3(b) showing that the least squares error does not change for fast relaxation times below ~10 fs. This value is an approximate upper limit of the fast relaxation time constant. Knowing that this value is that fast we can simplify Eq. (4) to:

$$e_{\text{LSMC,fit}}(t) \propto e_{shift}(t) + p[e_{shift}(t)] * \left[ H(t)\, t \exp\left(-t/\tau_{\text{a,slow}}\right) \right], \tag{5}$$

with $p$ being a proportionality factor. The result of fitting $e_{\text{LSMC,fit}}$ to $e_{\text{LSMC}}$ using Eq. (5) is shown in Fig. 4 for two carrier densities. An excellent agreement between both curves is obtained for slow anisotropy relaxation times of 120 fs and 110 fs for low and high carrier densities, respectively. The inset of Fig. 4 shows the dependences of the slow anisotropy relaxation times versus carrier density for excitation photon energies of 1.59 eV and 1.52 eV extracted from fits similar to the ones shown in Fig. 4(a) and (b). The data suggests that the anisotropy relaxation is faster for larger excess energy but additional studies are necessary to confirm this dependence. We comment on the density dependence of the slow anisotropy relaxation time further below.

In previous time-integrated experiments the $j_{\text{LSMC}}$ was mainly linked to the conduction band[24,27] because of the higher mobility and energy of the electrons as compared to holes. Yet, the band bending at the surface of our intrinsic GaAs material accelerates the holes towards the surface and the electrons away from the surface.[34] We believe that this effect enhances the hole contribution to the $j_{\text{LSMC}}$ such that two anisotropy relaxation times are found in our study. Since it is well known that typical scattering rates in the valence band are considerably larger than in the conduction band,[35] we assign the slow and fast relaxation times to the electrons and holes, respectively. We do not believe that the double-exponential decay of the anisotropy results from one single band corresponding to relaxation dynamics beyond the relaxation time approximation.

Finally we compare our measurements with previous anisotropy relaxation studies in which the polarization rotation of optical probe pulses has been measured time-resolved. Oudar et al.[25] found anisotropy relaxation times of ~200 fs in bulk GaAs at 77 K, a value which was close to the optical pulse width. Portella et al.[26] did a similar study with ~10 fs long probe pulses in bulk GaAs at room temperature and reported anisotropy relaxation times of few tens of femtoseconds with a density dependence of $n^{1/3}$ at $n \geq 10^{17} cm^{-3}$ suggesting that the anisotropy relaxation is mainly caused by carrier-carrier scattering. The anisotropy relaxation times reported in Ref. [26] are right in the middle between the fast and slow relaxation times obtained in our study providing. However, we do not observe a clear dependence of the slow anisotropy relaxation time on carrier density, but the quantitative values for this relaxation time agree very well with carrier-phonon scattering rates.[33] This suggests that carrier-phonon-scattering is the main mechanism of anisotropy relaxation in the conduction band for the carrier densities employed in this study. This finding is in line with previous time-integrated experiments on $j_{LSMC}$.[23] Moreover, our result does not contradict numerical simulations[20] in which anisotropy relaxation times in the conduction band of several 100 fs are obtained, yet, without considering carrier-phonon scattering. We believe that a more detailed comparison between our and previous anisotropy relaxation studies[25,26] is difficult due to two reasons. First, as already indicated in Ref. [20] optical probing of anisotropy might lead to relaxation times which depend on the spectral components of the probe beam. Second, it is not absolutely clear if screening of the surface field by optically induced carriers, which might also change the polarization of a probe beam,[36] has contributed to the results of Refs. [25,26]. In contrast, our probe of anisotropy relaxation is not influenced by these effects.

In conclusion, we demonstrated that the $j_{LSMC}$ can be used as a novel intrinsic probe of anisotropy relaxation in bulk GaAs. In general this method is applicable to all non-centrosymmetric materials. We observed a double-exponential decay or the optically induced anisotropy with the fast time constant being on the order of 10 fs and the slow time constant varying between 140 fs and 100 fs for carrier densities between $2.5 \times 10^{15}$ cm$^{-3}$ and $5 \times 10^{17}$ cm$^{-3}$. We attribute the fast and slow time constants to the valence and conduction band, respectively. The small dependence of the slow time constant on carrier density suggests that carrier-phonon scattering is the main origin of anisotropy relaxation in the conduction band.

We thank R. Podzimski and T. Meier for discussions, H. Marx and B. Hacke for technical assistance, and the Deutsche Forschungsgemeinschaft for financial support.

## References

[1] B.P. Zakharchenya, D.N. Mirlin, V.I. Perel', and I.I. Reshina, "Spectrum and Polarization of Hot-Electron Photoluminescence in Semiconductors," Sov. Phys. Uspekhi **25**, 143 (1982).

[2] V. V. Bel'kov, S.D. Ganichev, E.L. Ivchenko, S.A. Tarasenko, W. Weber, S. Giglberger, M. Olteanu, H.-P. Tranitz, S.N. Danilov, P. Schneider, W. Wegscheider, D. Weiss, and W. Prettl, "Magneto-Gyrotropic Photogalvanic Effects in Semiconductor Quantum Wells," J. Phys. Condens. Matter **17**, 3405 (2005).

[3] A.A. Bakun, B.P. Zakharchenya, A.A. Rogachev, M.N. Tkachuk, and V.G. Fleischer, "Observation of a Surface Photocurrent Caused by Optical Orientation of Electrons in a Semiconductor," JETP Lett. **40**, 1293 (1984).

[4] M. Bieler, K. Pierz, and U. Siegner, "Simultaneous Generation of Shift and Injection Currents in (110)-Grown GaAs/AlGaAs Quantum Wells," J. Appl. Phys. **100**, 083710 (2006).

[5] R.M. Elasticity and A. Escape, "Injection and Detection of a Spin-Polarized Current in a Light-Emitting Diode," Nature **402**, 787 (1999).

[6] E.L. Ivchenko, Y.B. Lyanda-Geller, and G.E. Pikus, "Magneto-Photogalvanic Effects in Noncentrosymmetric Crystals," Ferroelectrics **83**, 19 (1988).

[7] S. Priyadarshi, A.M. Racu, K. Pierz, U. Siegner, M. Bieler, H.T. Duc, J. Förstner, and T. Meier, "Reversal of Coherently Controlled Ultrafast Photocurrents by Band Mixing in Undoped GaAs Quantum Wells," Phys. Rev. Lett. **104**, 217401 (2010).

[8] J.E. Sipe and A.I. Shkrebtii, "Second-Order Optical Response in Semiconductors," Phys. Rev. B **61**, 5337 (2000).

[9] S.D. Ganichev, V. V. Bel'kov, S. a. Tarasenko, S.N. Danilov, S. Giglberger, C. Hoffmann, E.L. Ivchenko, D. Weiss, W. Wegscheider, C. Gerl, D. Schuh, J. Stahl, J. De Boeck, G. Borghs, and W. Prettl, "Zero-Bias Spin Separation," Nat. Phys. **2**, 609 (2006).

[10] S. Priyadarshi, K. Pierz, and M. Bieler, "Detection of the Anomalous Velocity with Subpicosecond Time Resolution in Semiconductor Nanostructures," Phys. Rev. Lett. **115**, 257401 (2015).

[11] C. Drexler, S. a. Tarasenko, P. Olbrich, J. Karch, M. Hirmer, F. Müller, M. Gmitra, J. Fabian, R. Yakimova, S. Lara-Avila, S. Kubatkin, M. Wang, R. Vajtai, P.M. Ajayan, J. Kono, and S.D. Ganichev, "Magnetic Quantum Ratchet Effect in Graphene.," Nat. Nanotechnol. **8**, 104 (2013).

[12] S.D. Ganichev, P. Schneider, V. V. Bel'kov, E.L. Ivchenko, S. Tarasenko, W. Wegscheider, D. Weiss, D. Schuh, B. Murdin, P. Phillips, C. Pidgeon, D. Clarke, M. Merrick, P. Murzyn, E. Beregulin, and W. Prettl, "Spin-Galvanic Effect due to Optical Spin Orientation in N-Type GaAs Quantum Well Structures," Phys. Rev. B **68**, 081302 (2003).

[13] R.D.R. Bhat, F. Nastos, A. Najmaie, and J.E. Sipe, "Pure Spin Current from One-Photon Absorption of Linearly Polarized Light in Noncentrosymmetric Semiconductors," Phys. Rev.

Lett. **94**, 096603 (2005).

[14] E.L. Ivchenko and S. a. Tarasenko, "Pure Spin Photocurrents," Semicond. Sci. Technol. **23**, 114007 (2008).

[15] T. Jungwirth, J. Wunderlich, and K. Olejník, "Spin Hall Effect Devices," Nat. Mater. **11**, 382 (2012).

[16] N. Locatelli, V. Cros, and J. Grollier, "Spin-Torque Building Blocks," Nat. Mater. **13**, 11 (2013).

[17] R. Jansen, "Silicon Spintronics," Nat. Mater. **11**, 400 (2012).

[18] P. Krzysteczko, J. Münchenberger, M. Schäfers, G. Reiss, and A. Thomas, "The Memristive Magnetic Tunnel Junction as a Nanoscopic Synapse-Neuron System," Adv. Mater. **24**, 762 (2012).

[19] R.L. Stamps, S. Breitkreutz, J. Åkerman, A. V Chumak, Y. Otani, G.E.W. Bauer, J.-U. Thiele, M. Bowen, S.A. Majetich, M. Kläui, I.L. Prejbeanu, B. Dieny, N.M. Dempsey, and B. Hillebrands, "The 2014 Magnetism Roadmap," J. Phys. D. Appl. Phys. **47**, 333001 (2014).

[20] R. Binder, H.S. Köhler, M. Bonitz, and N. Kwong, "Green's Function Description of Momentum-Orientation Relaxationof Photoexcited Electron Plasmas in Semiconductors," Phys. Rev. B **55**, 5110 (1997).

[21] C.B. Schmidt, S. Priyadarshi, S.A. Tarasenko, and M. Bieler, "Ultrafast Magneto-Photocurrents in GaAs: Separation of Surface and Bulk Contributions," Appl. Phys. Lett. **106**, 142108 (2015).

[22] A. V. Andrianov, E.V. V. Beregulin, Y.B. Lyanda-Geller, I.D.D. Yaroshetskii, and Y.B. Layanda-Geller, "Magnetic-Field-Induced Photovoltaic Effect in P-Type Gallium Arsenide," Sov. Physics, JETP **75**, 921 (1992).

[23] V.L. Alperovich, V.I. Belinicher, V.N. Novikov, and A.S. Terekhov, "Photogalvanic Effects Investigation in Gallium Arsenide," Ferroelectrics **45**, 1 (1982).

[24] V.L. Alperovich, A.O. Minaev, and A.S. Terekhov, "Ballistic Electron Transport through Epitaxial GaAs Films in a Magnetically Induced Surface Photocurrent," JETP Lett. **49**, 702 (1989).

[25] J.L. Oudar, A. Migus, D. Hulin, G. Grillon, J. Etchepare, and A. Antonetti, "Femtosecond Orientational Relaxation of Photoexcited Carriers in GaAs," Phys. Rev. Lett. **53**, 384 (1984).

[26] M.T. Portella, J.-Y. Bigot, R.W. Schoenlein, J.E. Cunningham, and C. V. Shank, "K-Space Carrier Dynamics in GaAs," Appl. Phys. Lett. **60**, 2123 (1992).

[27] V.L. Alperovich, V.I. Belinicher, V.N. Novikov, and A.S. Terekhov, "Surface Photovoltaic Effect in Solids . Theory and Experiment for Interband Transitions in Gallium Arsenide," Sov. Physics, JETP **53**, 1201 (1981).

[28] S. Priyadarshi, K. Pierz, and M. Bieler, in *SPIE OPTO*, edited by M. Betz, A.Y. Elezzabi,

J.-J. Song, and K.-T. Tsen (2013), p. 86231B.

[29] D. Côté, J.E. Sipe, and H.M. van Driel, "Simple Method for Calculating the Propagation of Terahertz Radiation in Experimental Geometries," J. Opt. Soc. Am. B **20**, 1374 (2003).

[30] A. Tomasino, A. Parisi, S. Stivala, P. Livreri, A.C. Cino, A.C. Busacca, M. Peccianti, and R. Morandotti, "Wideband THz Time Domain Spectroscopy Based on Optical Rectification and Electro-Optic Sampling," Sci. Rep. **3**, 3116 (2013).

[31] D. Côté, N. Laman, and H.M. van Driel, "Rectification and Shift Currents in GaAs," Appl. Phys. Lett. **80**, 905 (2002).

[32] M. Bieler, "THz Generation From Resonant Excitation of Semiconductor Nanostructures: Investigation of Second-Order Nonlinear Optical Effects," IEEE J. Sel. Top. Quantum Electron. **14**, 458 (2008).

[33] D. Snoke, "Density Dependence of Electron Scattering at Low Density," Phys. Rev. B **50**, 11583 (1994).

[34] T. Dekorsy, T. Pfeifer, W. Kütt, and H. Kurz, "Subpicosecond Carrier Transport in GaAs Surface-Space-Charge Fields," Phys. Rev. B **47**, 3842 (1993).

[35] J. Shah, *Ultrafast Spectroscopy of Semiconductros and Semiconductor Nanostructures*, 2nd ed. (Springer-Verlag Berlin Heidelberg New York, 1999).

[36] M. Bieler, K. Pierz, and U. Siegner, "Simultaneous Generation and Detection of Ultrashort Voltage Pulses in Low-Temperature Grown GaAs with below-Bandgap Laser Pulses," Appl. Phys. Lett. **94**, 051108 (2009).

**Figure 1:**

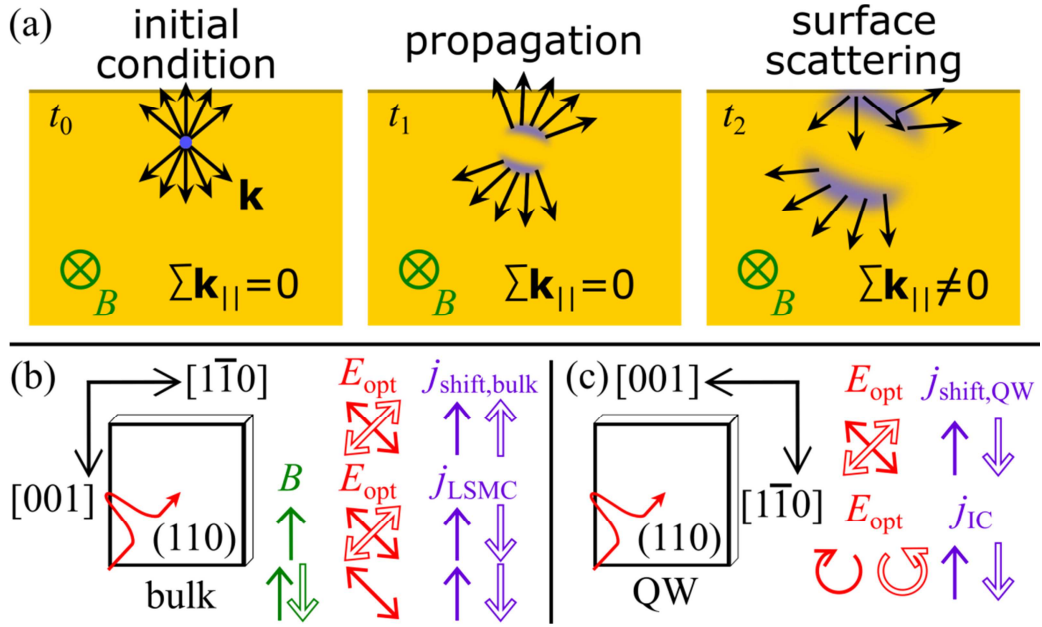

Fig. 1: (a) Generation of the $j_{LSMC}$ as a three-step process ($t_0$, $t_1$, and $t_2$) exemplarily shown for electrons. The momentum distribution of electrons is denoted by the solid arrows. The term $\sum \mathbf{k}_\parallel$ expresses the net in-plane momentum. (b) Dependence of the $j_{LSMC}$ and $j_{SC}$ on magnetic field and optical polarization in bulk GaAs. For the shown configuration the $j_{LSMC}$ flows parallel to $B$. (c) Dependence of the $j_{SC}$ and $j_{IC}$ on optical polarization in GaAs QWs.

**Figure 2:**

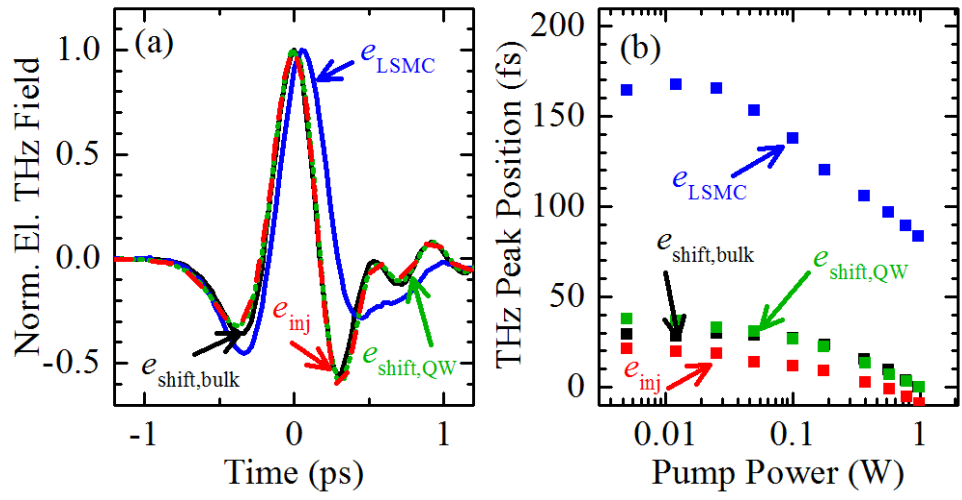

Fig. 2: (a) Normalized $e_{LSMC}$ and $e_{SC}$ in bulk GaAs and $e_{SC}$ and $e_{IC}$ in GaAs QWs. (b) Temporal peak position of the THz traces versus optical pump power.

**Figure 3:**

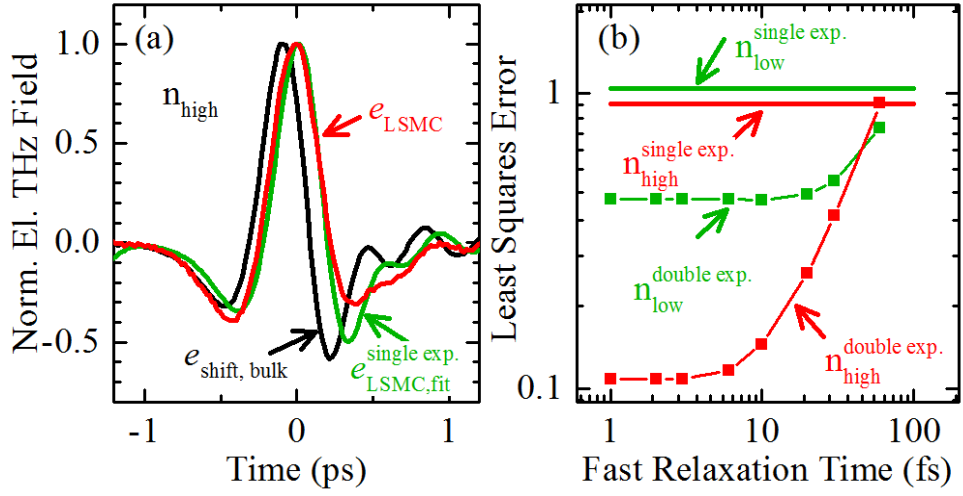

Fig. 3: (a) Normalized $e_{LSMC}$ (red), $e_{shift,bulk}$ (black), and $e_{shift,bulk}$ convolved with a single-exponentially decaying fit function (green) at 1.52 eV and a carrier density of $n_{high} = 5 \times 10^{17} \text{cm}^{-3}$. The parameters of the fit function were varied to minimize the difference between the red and green curves. (b) Least squares error employing a single exponentially decaying fit function (horizontal lines) and a double-exponentially decaying fit function versus time constant of the fast exponential decay for two different carrier densities ($n_{high}$ as above and $n_{low} = 2 \times 10^{16} \text{cm}^{-3}$).

**Figure 4:**

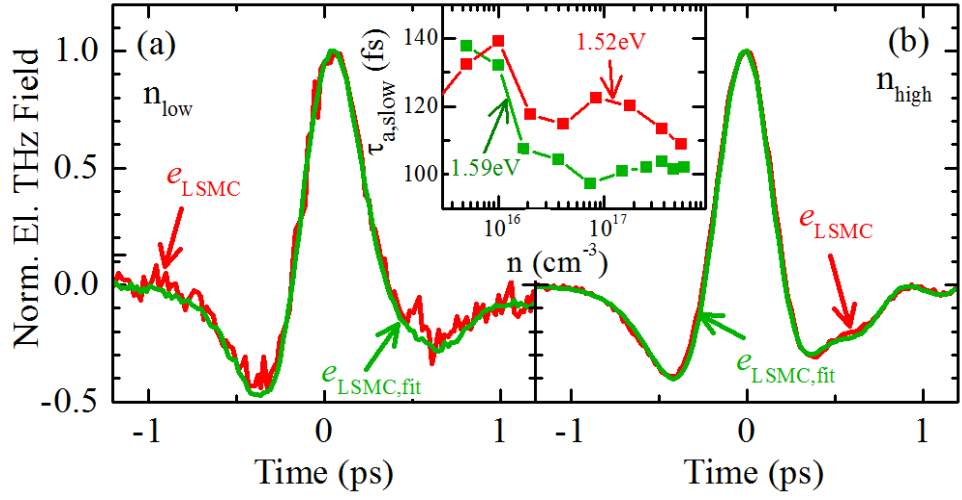

Fig. 4: Normalized $e_{\text{LSMC}}(t)$ and $e_{\text{LSMC,fit}}(t)$ using Eq. (5) at 1.52eV and for two carrier densities: (a) $n_{\text{low}} = 2 \times 10^{16} \text{cm}^{-3}$, (b) $n_{\text{high}} = 5 \times 10^{17} \text{cm}^{-3}$. Inset: Extracted slow anisotropy relaxation time $\tau_{a,\text{slow}}$ for 1.52 eV and 1.59 eV.